\documentclass[pra,aps,showpacs,showkeys,reprint]{revtex4}
\usepackage{graphicx}
\usepackage{amsmath}

\begin{document}
\title{\bf Generalized uncertainty principle and thermostatistics: a semiclassical approach}
\author{Mohammad Abbasiyan-Motlaq}
\author{Pouria Pedram}
\email[Electronic address: ]{p.pedram@srbiau.ac.ir}
\affiliation{Department of Physics, Science and Research Branch,
Islamic Azad University, Tehran, Iran}
\date{\today}

\begin{abstract}
We present an exact treatment of the thermodynamics of physical systems in the framework of
the generalized uncertainty principle (GUP). Our purpose is to study and compare the
consequences of two GUPs that one implies a minimal length while the
other predicts a minimal length and a maximal momentum. Using a
semiclassical method, we exactly calculate the modified internal
energies and heat capacities in the presence of generalized
commutation relations. We show that the total shift in these
quantities only depends on the deformed algebra not on the system
under study. Finally, the modified internal energy for an specific
physical system such as ideal gas is obtained in the framework of two different GUPs.
\end{abstract}


\maketitle

\section{Introduction}
As a significant consequence of various candidates of quantum
gravity such as string theory, loop quantum gravity, doubly special
relativity and black hole physics, the existence of a minimal
observable length and/or a maximal observable momentum is suggested
in the literature \cite{felder,1,2}. This minimal length
is of the order of the Planck length $\ell_{\mathrm{P}} =
\sqrt{{G\hbar}/{c^{3}}}\approx 10^{-35}m$, where $G$ is Newton's
gravitational constant.

Based on these theories, the Heisenberg uncertainty principle is
modified to the so-called Generalized (Gravitational) Uncertainty
Principle (GUP) \cite{9,10}. Due to this
modification, the Hamiltonian of the physical systems will be
modified which results in the deformation of the physical properties
of these systems in both quantum mechanical and classical levels. In the
context of the GUP framework, various problems such as harmonic
oscillator, hydrogen atom, gravitational quantum well, Casimir
effect, Landau levels, Lamb's shift, and particles scattering have
been investigated exactly or approximately in
Refs.~\cite{19,20,23,24,26,29,30,31,33}.

In the context of the statistical mechanics, Fityo developed a
semiclassical method for partition function evaluation based on
modification of elementary cells of phase space according to
deformed commutation relations and investigated the thermodynamical
properties of some physical systems up to the first order of the GUP
parameter \cite{34}. Moreover, using exact solutions of the
generalized Schr\"odinger equation \cite{12,18,24,35,36}, the
effects of the minimal length on partition function, internal
energy, and heat capacity in classical and quantum mechanical domains have been
studied numerically in Refs.~\cite{24,36,37,38,39}.

In this paper, we investigate the thermodynamical properties of the
physical systems in the context of the generalized uncertainty
principle. By using the  results of
Ref.~\cite{34}, we obtained   general relations that exactly calculate the
modified internal energies and heat capacities in the presence of
deformed commutation relations. Using exact solutions, we show that
the total shift in internal energies and heat capacities does not
depend on the physical systems and indeed it is related to the
deformed algebra. We have also compared the results of our calculations in two  different GUP frameworks. Finally, we obtain the modified internal energy
for the ideal gas in these GUP frameworks.

\section{The Generalized Uncertainty Principle}
First consider the following GUP in agreement with various theories of
quantum gravity \cite{3,4,5,6,7,8} which is proposed by Kempf, Mangano and
Mann (KMM) in one dimension \cite{11,12,13,14,15,16,17}
\begin{equation}\label{001}
\Delta X  \Delta P \geq \frac{ \hbar }{2} ( 1+ \beta  (\Delta P)^2 + \beta
\langle P \rangle^{2} ),
\end{equation}
where $ \beta = \beta _{0} / (c ^{2} M^{2}_{{\mathrm{Pl}}}) $ is
the GUP parameter, $ M_{{\mathrm{Pl}}} = \sqrt{\hbar c / G}$ is the
Planck mass and $ \beta _{0} $ is a dimensionless parameter of the
order of unity. The relation (\ref{001}) implies a minimal
observable length proportional to the Planck length $ ( \Delta X )
_{min} = \hbar \sqrt{ \beta } = \sqrt{ \beta _{0} }
\ell_{\mathrm{Pl}} $. The above generalized uncertainty relation
leads to modification of the canonical communication relations. In
three-dimensions we have \cite{11,12,13}
\begin{equation}\label{002}
\begin{array}{l}
[X_i,P_j]= i \hbar \left( ( 1 + \beta P^2 ) \delta_{ij} + \beta' P_i P_j \right),\\
\left[P_i,P_j\right]=0,\\
\left[X_i,X_j\right]= i \hbar \dfrac{2 \beta - \beta' + ( 2 \beta +
\beta' ) \beta P^{2}}{ 1 + \beta P^{2} } (P_{i} X_{j} - P_{j} X_{i}
),
\end{array}
\end{equation}
where $ i,j = 1,2,3 $ and $ \beta , \beta' $ are the GUP parameters.
For this case, the  minimal  length becomes $ (\Delta X)_{min} =
\hbar \sqrt{ \beta + \beta' } $.

To incorporate the idea of maximal observable momentum in agreement
with doubly special relativity theories, we use the recently proposed generalized uncertainty principle (GUP$^*$) which
implies both the minimal length uncertainty and maximal observable
momentum \cite{14,15}
\begin{eqnarray}\label{005}
\begin{array}{l}
\left[ X_{i} , P_{j}\right]  = \dfrac{ i \hbar \delta _{ij} }{ 1 - \beta P^{2} },\\
\left[P_{i} , P_{j}\right]  = 0,\\
\left[ X_{i} , X_{j}\right] = \dfrac{2 i \hbar \beta }{(1 - \beta
P^{2} )^{2}} ( P_{i} X_{j} - P_{j} X_{i} ),
\end{array}
\end{eqnarray}
where $(\Delta X)_{min}=\frac{3\sqrt{3}}{4}\hbar\sqrt{\beta}$ and
the momentum of the particle cannot exceed $ {1}/{ \sqrt{ \beta } }
$, i.e., $P_{max}={1}/{ \sqrt{ \beta } }$. Note that, these generalized
commutation relations are the particular forms of the general
relations \cite{32}
\begin{eqnarray}\label{012}
\begin{array}{l}
\left[ X_{i} , P_{j}\right] = i \hbar\, f_{ij} (\beta ,\beta',P),\\
\left[P_{i} , P_{j}\right]  = i \hbar\, h_{ij} (\beta,\beta',P),\\
\left[ X_{i} , X_{j}\right] = i \hbar\, g_{ij} (\beta, \beta',P).
 \end{array}
\end{eqnarray}
Here, $ \{\beta, \beta'\} $ are the GUP parameters and $ \{f_{ij} ,
g_{ij} , h_{ij}\} $ are deformation   functions. In the limit $
\beta , \beta' \rightarrow 0 $,  $ f_{ij} $ goes to unity while $
g_{ij} $ and $ h_{ij} $ go to zero. Thus, the position and momentum
operators tend to the ordinary position and momentum operators $
x_{i} $ and $ p_{i} $ satisfying $ [ x_{i} , p_{j} ] = i \hbar
\delta _{ij} $. In the classical limit $ \hbar \rightarrow 0 $ the
quantum commutation relations lead to the Poisson brackets as
\begin{equation}\label{015}
\dfrac{1}{i\hbar} [A,B]\Rightarrow \lbrace A,B \rbrace.
\end{equation}
The effects of the GUP on classical and quantum mechanical systems are addressed in
Refs.~\cite{felder,18,21,22,25,28,32,27}.

\section{GUP and the modified thermodynamics}
The partition function for a system with $ N $ non-interacting  particles in  the GUP
framework is \cite{34}
\begin{equation}\label{025}
Z_{N} = \dfrac{1}{h^{3N}} \int  \dfrac{\mathrm{d}^{3N} X \mathrm{d}^{3N} P}{J}\exp \left[-\dfrac{H(X,P)}{k_{B}T} \right] ,
\end{equation}
where $ H(X,P) $ is the Hamiltonian of the system, $ k_{B} $ is
Boltzman's constant, $ T $ is the  temperature, and $
J=\frac{\partial (X_1,P_1,\dots,X_D,P_D)}{\partial
(x_1,p_1,\dots,x_D,p_D)} $ is the Jacobian of the transformation in
$D$-dimensions. Indeed, the Jacobian can be read off from the
modified poisson brackets. In three-dimensions we have \cite{34}
\begin{eqnarray}
&&\frac{\partial (X_1,P_1,X_2,P_2,X_3,P_3)}{\partial
(x_1,p_1,x_2,p_2,x_3,p_3)}=
\left\{X_1,P_1\right\}\left\{X_2,P_2\right\}\left\{X_3,P_3\right\}-
\left\{X_1,P_3\right\}\left\{P_1,P_2\right\}\left\{X_2,X_3\right\}-\nonumber\\
&&\left\{X_1,P_2\right\}\left\{X_2,P_1\right\}\left\{X_3,P_3\right\}-
\left\{X_1,P_3\right\}\left\{X_2,P_2\right\}\left\{X_3,P_1\right\}-
\left\{X_1,P_1\right\}\left\{X_2,P_3\right\}\left\{X_3,P_2\right\}+\nonumber\\
&&\left\{X_1,X_2\right\}\left\{P_1,P_3\right\}\left\{X_3,P_2\right\}+
\left\{X_1,P_3\right\}\left\{X_2,P_1\right\}\left\{X_3,P_2\right\}-
\left\{X_1,X_2\right\}\left\{P_2,P_3\right\}\left\{X_3,P_1\right\}+\nonumber\\
&&\left\{X_1,P_2\right\}\left\{X_2,X_3\right\}\left\{P_1,P_3\right\}-
\left\{X_1,X_3\right\}\left\{P_1,P_3\right\}\left\{X_2,P_2\right\}+
\left\{X_1,X_3\right\}\left\{X_2,P_1\right\}\left\{P_2,P_3\right\}+\nonumber\\
&&\left\{X_1,X_3\right\}\left\{P_1,P_2\right\}\left\{X_2,P_3\right\}-
\left\{X_1,X_2\right\}\left\{P_1,P_2\right\}\left\{X_3,P_3\right\}-
\left\{X_1,P_1\right\}\left\{X_2,X_3\right\}\left\{P_2,P_3\right\}+\nonumber\\
&&\left\{X_1,P_2\right\}\left\{X_2,P_3\right\}\left\{X_3,P_1\right\}.
\end{eqnarray}

For the deformed commutation relations (\ref{012}), the Hamiltonian
can be written in the following form
\begin{equation}\label{026}
H = \dfrac{P^{2}}{2m} + U(X),
\end{equation}
and the relation (\ref{025}) becomes
\begin{equation}\label{027}
Z_{N} = \dfrac{1}{h^{3N}} \displaystyle { \int \mathrm{d}^{3N} X \exp \Big[-\dfrac{U(X)}{k_{B}T} \Big] } \displaystyle { \int \mathrm{d}^{3N} P \dfrac{\exp \Big[ - \dfrac{P^{2}}{2mk_{B}T} \Big] } { J ( \beta , \beta' , P ) } },
\end{equation}
where $ U(X) $ is the potential function.
In ordinary thermodynamics we have
\begin{equation}\label{028}
Z^{0}_{N} = \dfrac{1}{h^{3N}} \displaystyle { \int \mathrm{d}^{3N}x \, \mathrm{d}^{3N}p \, \exp \Big[ -\dfrac{H_{0}(x,p)}{k_{B}T} \Big] },
\end{equation}
where $ H_{0} = {p^{2}}/{2m} + U(x) $ is the  Hamiltonian for the
non-deformed case. Now, since
\begin{equation}\label{029}
{ \int \mathrm{d}^{3N}x \, \exp \Big[ -\dfrac{U(x)}{k_{B}T} \Big] }
= \hbar ^{3N} Z_{N}^{0} \left( 2 \pi m k_{B} T
\right)^{-\frac{3}{2}N},
\end{equation}
we obtain
\begin{equation}\label{030}
Z_{N} = Z^{0}_{N}  \left( 2 \pi m k_{B} T \right)^{-\frac{3}{2}N} \displaystyle { \int \mathrm{d}^{3N} P \dfrac{\exp \Big[ - \dfrac{P^{2}}{2mk_{B}T} \Big] } { J ( \beta , \beta' , P ) } }.
\end{equation}
Thus, the modified internal energy $ E = - \frac{ \partial } { \partial (k_{B}T) } \ln Z $ and heat capacity $ C = \frac{ \partial E } { \partial T } $, are given by
\begin{equation}\label{031}
\begin{array}{l}
E = E^{0} - \dfrac{3}{2} Nk_{B}T + \dfrac{N}{2m} \dfrac{S_{2}}{S_{0}},\\
C = C^{0} - \dfrac{3}{2} Nk_{B} - \dfrac{N}{4m^{2}k_{B}T^{2}}\dfrac{S_{4}S_{0}-(S_{2})^2}{(S_{0})^2},
\end{array}
\end{equation}
where $ S_{n}= \displaystyle { \int \mathrm{d}^{3} P \, P^{n}
\dfrac{\exp \Big[ - \dfrac{P^{2}}{2mk_{B}T} \Big] } { J ( \beta ,
\beta' , P ) } }  $. Here, $ E^{0} $ and $ C^{0} $ are  the internal
energy and the heat capacity in ordinary  thermodynamics,
respectively, i.e., $ E^{0} = - \frac{ \partial } { \partial
(k_{B}T) } \ln Z^{0} $ and $ C^{0} = \frac{ \partial E^{0} } {
\partial T } $. After integrating out the angular parts, the above
equations can be written as
\begin{equation}\label{033}
\begin{array}{l}
E = E^{0} - \dfrac{3}{2}Nk_{B}T + \dfrac{N}{2m} \dfrac{s_{4}}{s_{2}},\\
C = C^{0} - \dfrac{3}{2} Nk_{B} - \dfrac{N}{4m^{2}k_{B}T^{2}}\dfrac{s_{6}s_{2}-(s_{4})^2}{(s_{2})^2},
\end{array}
\end{equation}
where $ s_{n}= \displaystyle { \int \mathrm{d}P \, P^{n}
\dfrac{\exp \Big[ - \dfrac{P^{2}}{2mk_{B}T} \Big] } { J ( \beta ,
\beta' , P ) } }  $. Other modified thermodynamical quantities such
as Helmholtz free energy $ A = - k_{B} T \ln Z $ and entropy $ S =
k_{B} \ln Z + {E}/{T} $ can be obtained in a similar manner.

\section{Applications}
In the following subsections, we investigate the effects of two GUPs
(\ref{002},\ref{005}) on the thermodynamical properties of the
physical systems. The KMM's GUP framework implies a minimal length
while the high order GUP framework (GUP$^*$) predicts minimal length
and maximal momentum. We also compare the results for both GUPs.

\subsection{KMM's GUP: minimal length}
In three-dimensional space, the KMM's GUP is given by
Eq.~(\ref{002}). For this case, the Jacobian of the transformation
reads \cite{34}
\begin{equation}\label{038}
J = (1 + \beta P^{2})^{2} (1 + (\beta + \beta') P^{2}).
\end{equation}
So, using Eq.~(\ref{033}) we obtain the following exact relation for the internal energy
\begin{equation}\label{039}
E = E^{0} - \dfrac{3}{2} Nk_{B}T + \dfrac{N}{2m \beta }\dfrac{ -\Omega_1
\Gamma \Big( \dfrac{1}{2} , \dfrac{1}{2 \beta \xi} \Big) +
\sqrt{\beta} \bigg( \Omega_2 + \Omega_3 \Gamma \Big(\dfrac{1}{2} , \dfrac{1}{2 (\beta + \beta') \xi}\Big ) \bigg) }{ \Omega_4 \Gamma \Big( \dfrac{1}{2} , \dfrac{1}{2 \beta \xi} \Big) +
 \Omega_5- \Omega_6 \Gamma \Big(\dfrac{1}{2} , \dfrac{1}{2 (\beta + \beta') \xi}\Big ) },
\end{equation}
where $\xi= mk_{B}T$ and
\begin{eqnarray}\label{040}
\left\{
\begin{array}{l}
\Omega_1=\sqrt{ \beta' + \beta } \Big( - \beta'  - \beta'\beta\xi +  2 \beta^{2} \xi\Big)\exp\left({ \frac{1}{2\beta \xi} }\right),\\
\Omega_2=-\beta' \sqrt{2 (\beta' + \beta) \xi },\\
\Omega_3=2 \beta^{2} \xi \exp\left({ \frac{ 1 }{ 2 (\beta + \beta') \xi } }\right), \\
\Omega_4=\sqrt{ \beta' + \beta } \Big(- \beta'  + \beta'\beta\xi +  2 \beta^{2} \xi \Big)\exp\left({ \frac{1}{2\beta \xi} }\right),\\
\Omega_5=\beta' \sqrt{2\beta (\beta' + \beta) \xi },\\
\Omega_6=2 \beta ( \beta + \beta' ) \sqrt{ \beta } \, \exp\left({ \frac{ 1 } { 2 (\beta + \beta') \xi }}\right).
\end{array}\right.
\end{eqnarray}
The exact heat capacity is given by
\begin{eqnarray}
C=&& C^{0} - \dfrac{3}{2} Nk_{B} +\dfrac{1}{2} N\sqrt{ \dfrac{ k_{B} }{ mT } }( \beta')^{2}
\nonumber\\
&& \times\dfrac{
 \Psi_1 \Gamma \Big(\dfrac{1}{2} , \big(\dfrac{1}{2 \beta\xi}\big)^{2} \Big)  +
\Psi_2 - \Psi_3 \Gamma\Big (\dfrac{1}{2} , \dfrac{1}{2 (\beta + \beta')\xi} \Big)  +
  \Psi_4 +
 \Psi_5\Gamma \Big(\dfrac{1}{2} , \dfrac{1}{2 (\beta + \beta') \xi}\Big )   }{ \sqrt{ \beta (\beta + \beta') } \bigg[ \Psi_6 \Gamma\Big (\dfrac{1}{2} , \dfrac{1}{2 \xi}\Big )  +
\xi \sqrt{ \beta } \Big[ \Psi_7 - \Psi_8 \Gamma \Big(\dfrac{1}{2} , \dfrac{1}{2 (\beta+\beta') \xi}\Big ) \Big] \bigg]^{2} },\label{040-1}
\end{eqnarray}
in which
\begin{eqnarray}\label{041}
\left\{
\begin{array}{l}
\Psi_1=-2\big ( \beta + \beta'\big )^{ \frac{3}{2} } \sqrt{\pi \beta \xi} \, \exp\left({ \frac{ 1 } {  \beta\xi }}\right),\\
\Psi_2=   2 \beta' \sqrt{\beta(\beta + \beta')\xi},\\
\Psi_3=\sqrt{2\beta}\Big ( \beta' + 2\beta \beta' \xi +
2  \beta^{2}\xi\Big) \exp\left({ \frac{ 1 } { 2 (\beta + \beta')\xi }}\right),\\
\Psi_4= \sqrt{2(\beta + \beta')} \Big(-\beta' + \beta \beta' \xi + 2 \beta^{2}\xi\Big)  \exp\left({ \frac{ 1 } { 2 \beta\xi }}\right)\Gamma \left(\dfrac{1}{2} , \dfrac{1}{2 \beta \xi}\right),\\
\Psi_5=\sqrt{ \dfrac{ 1 }{\xi} }\Big  (\beta' + 3 \beta \beta'\xi + 2 \beta^{2}\xi\Big) \exp\left({ \frac{ \beta' + 2\beta }{ 2\beta (\beta + \beta')\xi } }\right)\Gamma\left( (\dfrac{1}{2} , \dfrac{1}{2 \beta\xi}\right),\\
\Psi_6=\sqrt{ \beta + \beta'} \Big( -\beta' + \beta \beta' \xi + 2 \beta^{2}\xi \Big)\exp\left({ \frac{ 1 }{ 2\beta \xi } }\right),\\
\Psi_7=\beta' \sqrt{ \dfrac{ 2(\beta + \beta') }{ \xi } },\\
\Psi_8=2 \beta \Big(\beta + \beta'\Big) \exp\left({ \frac{ 1 }{ 2(\beta + \beta') \xi } }\right),
\end{array}\right.
\end{eqnarray}
and $ \Gamma (a,x) = \int _{x}^{\infty} t^{a-1} e^{-t} \mathrm{d}t
$ is the  incomplete gamma function. In the limit $ \beta ,
\beta' \longrightarrow 0 $, the internal energy and heat capacity
tend to $ E^{0} $ and $ C^{0} $,
respectively, as it is expected.

Notice  that, although the changes in the energies and heat
capacities depend on the modified algebra, these changes are similar
for all physical systems. It is worth mentioning that the effects of
GUP on various physical systems are also addressed in
Refs.~\cite{19,20,23,26,29,30,31,33}. In particular, the effects of
minimal length on thermostatistics of classical and quantum
mechanical systems have been studied in Refs.~\cite{36,37,38,39}.

\subsection{GUP$^*$: minimal length and maximal momentum}
For GUP$^*$ the Jacobian of transformation becomes
\begin{equation}\label{053}
J = \dfrac{ 1 }{ ( 1 - \beta P^{2} )^{3} },
\end{equation}
Now, using Eq.~(\ref{033}) we obtain the exact internal energy as
\begin{align}\label{054}
E = E^{0} -\frac{3}{2}Nk_BT+\frac{N}{2m}\frac{\Theta_1\exp\left( \dfrac{ -1 }{  2 \beta\xi  }\right) - \Theta_2\Gamma \left(\dfrac{1}{2} , \dfrac{1}{2 \beta\xi } \right)}{\Theta_3\exp\left( \dfrac{ -1 }{  2 \beta \xi  }\right)+ \Theta_4\Gamma \left(\dfrac{1}{2} , \dfrac{1}{2 \beta \xi } \right)},
\end{align}
where
\begin{eqnarray}\label{055}
\left\{
\begin{array}{l}
\Theta_1=3\sqrt{\beta}\xi T^{2}\Big(1-315\beta^2 \xi T\Big),\\
\Theta_2=\frac{3}{ \sqrt{2} }  T^{\frac{3}{2}} \Big(- \beta \xi + 15 \beta^{2}\xi^{2}  - 105\beta^{3}\xi^{3}  +  315 \beta^{4} \xi^{4}  \Big),\\
\Theta_3=\sqrt{ \beta }\xi T  \Big( 1 -10 \beta\xi  + 105  \beta^{2}\xi^{2}\Big ),\\
\Theta_4=\frac{ 1 }{ \sqrt{2} } \xi T^{ \frac{1}{2} } \Big( 1 - 9 \beta\xi  +45 \beta^{2}\xi^{2} - 105 \beta^{3} \xi^{3}  \Big).
\end{array}\right.
\end{eqnarray}
For the heat capacity, we exactly find
\begin{eqnarray}\label{055-1}
C &=& C^{0} - \dfrac{3}{2} Nk_{B} +3 Nk_{B}
\nonumber\\
&&\times\dfrac{ \Delta_1
-2\Delta_2 \exp\left( \dfrac{ 1 }{  2 \beta \xi  }\right)
\Gamma\left (\dfrac{1}{2} , \dfrac{1}{2 \beta \xi }\right ) +
\sqrt{\pi} \Delta_3 \exp\left( \dfrac{ 1 }{   \beta \xi  }\right)
\Gamma \left(\dfrac{1}{2} ,\Big (\dfrac{1}{2 \beta\xi }\Big)^{2}\right) }{\left[ \Delta_4 + \Delta_5 \exp\left( \dfrac{ 1 }{  2 \beta\xi }\right)
\Gamma\left (\dfrac{1}{2} , \dfrac{1}{2 \beta\xi }\right)\right]^{2}},
\end{eqnarray}
where
\begin{eqnarray}\label{056}
\left\{
\begin{array}{l}
\Delta_1=18 \beta\xi   + 210 T   \Big ( 8\beta^{3}\xi^{2} -60 \beta^{4} \xi^{3} + 315\beta^{5} \xi^{4}\Big ),\\
\Delta_2= \sqrt{ 2 \beta \xi }\Big(-5 + 11 \beta\xi  -630 \beta^{2}\xi^{2}  + 4410 \beta^{3}\xi^{3}
 -17325 \beta^{4} \xi^{4} + 33075 \beta^{5} \xi^{5}
\Big),\\
\Delta_3= 1 - 30 \beta\xi +405 \beta^{2}\xi^{2}  -2940 \beta^{3}\xi^{3}  +11655\beta^{4}\xi^{4}
 -28350 \beta^{5}\xi^{5}
 +33075 \beta^{6}\xi^{6},\\
\Delta_4=  \sqrt{ \beta\xi }\Big ( -2 +20 \beta\xi  -210\beta^{2} \xi^{2}\Big ),\\
\Delta_5=\sqrt{2}\Big ( -1 + 9 \beta\xi
 -45 \beta^{2}\xi^{2}  + 105 \beta^{3}\xi^{3} \Big).
 \end{array}\right.
\end{eqnarray}

\begin{figure}
  \begin{center}
  \includegraphics[width=10cm]{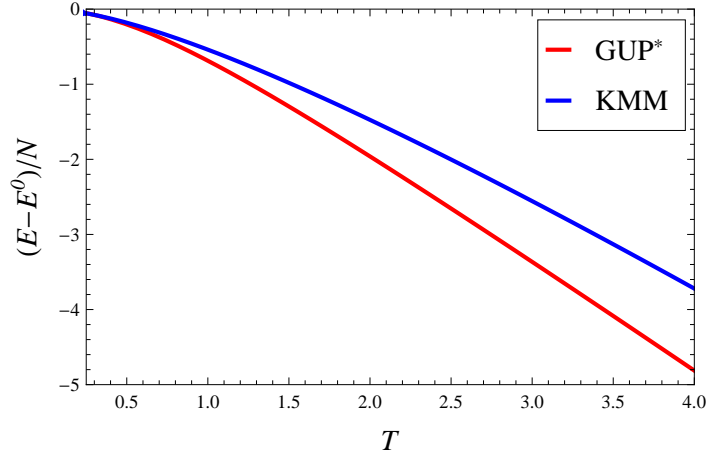}
  \caption{\label{fig1}The shift of internal energy  $ {(E-E^{0})}/{N} $  versus $ T$
  for KMM's GUP (blue line) and GUP$^*$ (red line). We set $ m=k_{B}=1 $ and $ \beta=\beta'=0.1 $. }
\end{center}
\end{figure}

\begin{figure}
  \begin{center}
  \includegraphics[width=10cm]{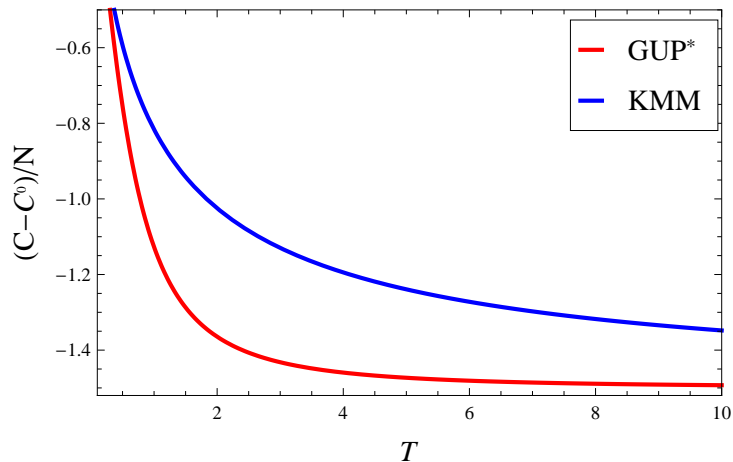}
\caption{\label{fig2}The shift of heat capacity  $ {(C-C^{0})}/{N} $
versus $T $ for KMM's GUP (blue line) and GUP$^*$ (red line). We
set $ m=k_{B}=1 $ and $ \beta=\beta'=0.1 $.}
\end{center}
\end{figure}

It is worth mentioning that the shift in the internal energies and
heat capacities for physical systems  does not depend on the type of
the system. For comparison, we plotted ${(E-E^{0})}/{N}$ (see
Fig.~\ref{fig1}) and ${(C-C^{0})}/{N}$  (see Fig.~\ref{fig2}) in
terms of temperature in KMM's GUP and GUP$^*$ frameworks. As it can
be seen from Fig.~\ref{fig1}, the energy shift of physical systems
for GUP$^*$ is greater than the KMM's GUP. Fig.~\ref{fig2} shows the
similar result for the heat capacity. Moreover, in the high
temperature limit, the heat capacity tends to zero for both cases,
namely $C(T\rightarrow\infty)=0$ in agreement with Ref.~\cite{34}.
Thus, according to the relation $ C = \frac{
\partial E } { \partial T } $, the internal energy asymptotically tends to a maximum value.

\begin{figure}
  \begin{center}
  \includegraphics[width=10cm]{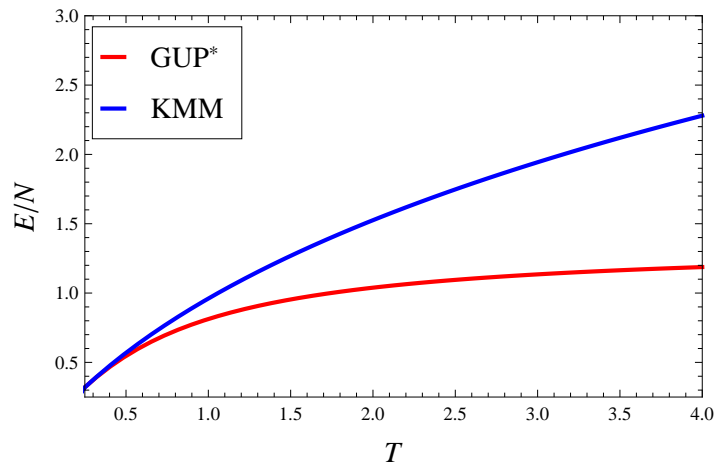}
\caption{\label{fig3}The internal energy of ideal gas $ {E}/{N}$
versus $ T $ for KMM's GUP (blue line) and GUP$^*$ (red line). We
set $ m=k_{B}=1 $ and $ \beta=\beta'=0.1 $. }
\end{center}
\end{figure}

Finally, we have shown the internal energy versus temperature for
the ideal gas system in Fig.~\ref{fig3}. As this figure shows, in
low temperature limit, the behavior of ideal gas's internal energy
for both the KMM's GUP and GUP$^*$ coincides. However, in high
temperature limit, the internal energy of the ideal gas in the KMM's
GUP is greater than the internal energy in the GUP$^*$ framework and contains a maximum value
in both scenarios. The exact maximum value  of  internal energy for the ideal gas in the KMM's GUP  is  given by
\begin{eqnarray}\label{098}
E_{max}^{KMM}= \dfrac{ 1+\dfrac{2\sqrt{\beta}}{\sqrt{\beta'+\beta}} }{ 2m\beta },
\end{eqnarray}
It is worthwhile to note that in the GUP$^*$ framework the momentum
of particle can not exceed $ {1}/{ \sqrt{ \beta } }$. Therefore, we
expect that the temperature of the system represent a maximum value.
At this temperature  the heat capacity of systems is zero. Using
Eq.~(\ref{033}) we obtain the following relation to calculate  the
value of maximum temperature for the ideal gas system
\begin{eqnarray}\label{099}
s_{6}s_{2}-(s_{4})^2 = 0,
\end{eqnarray}
which can be solved numerically. At high temperature limit, i.e.,
$2mk_{B}T\gg 1$, we have
\begin{eqnarray}\label{100}
T_{max}\approx \dfrac{7}{30mk_{B}\beta}.
\end{eqnarray}
At this temperature, $ E_{max}$ for the ideal gas becomes
\begin{eqnarray}\label{101}
E_{max}^{GUP^*}&=& \dfrac{ 63\Big(24010+1400mk_{B}-157\sqrt{105\pi}e^\frac{15}{7}(mk_{B})^\frac{3}{2}Erf[\sqrt{\frac{15}{7}}]\Big) }{20m^3k^2_{B}\beta \Big(55230+29e^\frac{15}{7}\sqrt{105\pi mk_{B}}Erf[\sqrt{\frac{15}{7}}]  \Big) }.
\end{eqnarray}
For instance, for $ m=k_{B}=1 $ and $ \beta=\beta'=0.1 $ these
maximum  values are found to be 12.07 and 1.07 for the KMM's GUP and
GUP$^*$, respectively. Also, the internal energy and heat capacity
of other  physical systems such as harmonic oscillator can be
calculated by substituting the value of $ E_{0}$ and $ C_{0}$ in
Eq.~(\ref{033}). Furthermore, the modified partition function
Eq.~(\ref{030}) up to the first order of GUP parameter in KMM's GUP
framework \cite{34} and GUP$^*$, are respectively given by
\begin{eqnarray}\label{103}
Z^{KMM}&=& Z_{0}\Big(1-3(3\beta+\beta')mk_{B}T+\cdot\cdot\cdot\Big),\\
Z^{GUP^*}&=&Z_{0} \Big(1-9\beta mk_{B}T+\cdot\cdot\cdot\Big).\label{104}
\end{eqnarray}
In fact, for $ \beta'=0 $, Eqs.~(\ref{103}) and (\ref{104}) become
identical. This is an expected result, due to equality of two GUP
frameworks to the first order of GUP parameter when $ \beta'=0$.

\section{Conclusions}
In this paper, we have studied the thermodynamics of physical
systems in the framework of the generalized uncertainty principle
\cite{felder,1,2,36,37,38,39}. We obtained  exact semiclassical
relations in order  to calculate the internal energy and heat capacity of
physical systems in a general deformed algebra. We showed that the
shift in internal energies and heat capacities are the same for all
physical systems and only depends on the chosen deformed algebra. We
applied this method for GUP$^*$ and KMM's GUP and obtained the
GUP-corrected thermodynamical variables. Furthermore, we
investigated the behavior of internal energy for the ideal gas
system in these frameworks. It is shown that for small temperature
limit, the behavior of the internal energy is the same for both
cases. However, in high temperature limit, the internal energy
monotonically increases versus temperature such that
$E_{\mathrm{KMM}}>E_{\mathrm{GUP^*}}$.

\end{document}